\documentclass[aps,twocolumn,superscriptaddress,showpacs]{revtex4}

\usepackage[dvips]{graphicx}  
\newcommand{\nc}{\newcommand}           
\nc{\vc}[1]     {\mbox{\boldmath $#1$}} 
\nc{\bra}       {\langle}               
\nc{\ket}       {\rangle}               
\nc{\bras}[1]   {\langle#1|}            
\nc{\kets}[1]   {|#1\rangle}            
\nc{\beq}     {\begin{eqnarray}}
\nc{\eeq}    {\end{eqnarray}}

\nc{\mydraft}	{\setlength{\topmargin}{-1.0cm}} 
\mydraft

\begin{document}

\title{Structures in $^{9,10}$Be and $^{10}$B studied with tensor-optimized shell model}

\author{Takayuki Myo\footnote{myo@ge.oit.ac.jp}}
\affiliation{General Education, Faculty of Engineering, Osaka Institute of Technology, Osaka, Osaka 535-8585, Japan}
\affiliation{Research Center for Nuclear Physics (RCNP), Osaka University, Ibaraki, Osaka 567-0047, Japan}

\author{Atsushi Umeya\footnote{aumeya@nit.ac.jp}}
\affiliation{Human Science and Common Education, Faculty of Engineering, Nippon Institute of Technology, Saitama 345-8501, Japan}

\author{Hiroshi Toki\footnote{toki@rcnp.osaka-u.ac.jp}}
\affiliation{Research Center for Nuclear Physics (RCNP), Osaka University, Ibaraki, Osaka 567-0047, Japan}

\author{Kiyomi Ikeda\footnote{k-ikeda@postman.riken.go.jp}}
\affiliation{RIKEN Nishina Center, Wako, Saitama 351-0198, Japan}

\date{\today}

\begin{abstract}%
We investigate the structures of $^{9,10}$Be and $^{10}$B with the tensor-optimized shell model (TOSM)
using the effective interaction based on the bare nucleon-nucleon interaction AV8$^\prime$.
The tensor correlation is treated in TOSM with the full optimization of 2p2h configurations including high-momentum components.
The short-range correlation is described in the unitary correlation operator method (UCOM). 
It is found that the level orders of the low-lying states of $^{9,10}$Be and $^{10}$B are entirely reproduced.
For $^9$Be, ground band states are located relatively in higher energy than the experiments, 
which indicates the missing $\alpha$ clustering correlation in these states as seen in the case of $^8$Be with TOSM.
In addition, the tensor force gives the larger attraction for $T$=1/2 states than for $T$=3/2 ones for $^9$Be.
For $^{10}$Be, the tensor contribution of $0^+_2$ shows the largest value among the $0^+$ states. This can be related to the $\alpha$ clustering correlation in this state.
It is also found that the level order of three nuclei depends on the tensor force in comparison with the results obtained with the Minnesota interaction without the tensor force.
\end{abstract}


\maketitle

\section{Introduction}

It is an important problem in nuclear physics to realize how the nucleon-nucleon ($NN$) interaction explains the nuclear structure.
The Green's function Monte Carlo method~(GFMC) recently makes it possible to describe nuclei up to a mass number around $A\sim 12$ with bare $NN$ interaction \cite{pieper01, pudliner97}. 
Toward the description of nuclei with larger mass number, it is important to develop a method of treating the system with large nucleon numbers using the bare interaction.

One of the characteristics of the $NN$ interaction is the strong tensor force which is mainly caused by the pion exchange and explains the large amount of the binding energy of nuclei \cite{pieper01,wiringa00}.
The $NN$ tensor force produces the strong $sd$ coupling in energy, which makes the deuteron bound. 
The $d$-wave component in the wave function is spatially compact because of the high momentum component induced by the tensor force \cite{ikeda10}. 
Experimentally, the possible evidence of the high-momentum component of a nucleon coming from the tensor correlation in nucleus has been developed using the ($p$,$d$) reaction~\cite{tanihata10,ong13}.

It is known that the tensor correlation in the $\alpha$ particle is generally strong~\cite{akaishi86,kamada01}.
The large binding energy of the $\alpha$ particle can be related to the presence of the tensor force.
In light nuclei, the $\alpha$ particles are often developed as a cluster, e.g., in $^8$Be and in $^{12}$C such as the Hoyle state \cite{ikeda68,horiuchi12}.
For $^8$Be, the $0^+_1$, $2^+_1$, and $4^+_1$ states are considered to be the two-$\alpha$ cluster states.
The excited states above $4^+_1$ can be regarded as shell-like states, because the $\alpha$ decay process is not often favored for these states.
For $^{12}$C, the ground state is mainly a shell-like state and some states in the excited states are considered to be the triple-$\alpha$ states including the Hoyle state.
It is worthy to investigate the mechanism of the formations of the shell-like state and the $\alpha$ cluster state.
It is interesting to investigate how the tensor force affects this phenomenon.

In the bare $NN$ interaction, the tensor force has the intermediate range character, so that we can include the tensor force contribution within a reasonable shell model space~\cite{myo07,myo09,myo11,myo12}.  
We call this method the tensor-optimized shell model (TOSM).  
In the TOSM, the shell model basis states are employed optimizing the two particle-two hole (2p2h) states fully. 
The particle states are not truncated, where spatial shrinkage of the particle states is necessary to describe the strong tensor correlation~\cite{toki02,sugimoto04,ogawa06}.
In addition, we use the unitary correlation operator method (UCOM) to describe the short-range correlation in the $NN$ interaction \cite{feldmeier98,roth10}. 
We describe the nuclei with TOSM+UCOM using a bare $NN$ interaction.  

Using TOSM, we have obtained successful results for He and Li isotopes and $^8$Be \cite{myo11,myo05,myo06,myo12}.
It is found that the $(p_{1/2})^2(s_{1/2})^{-2}$ type of the 2p2h configuration of the $pn$ pair is strongly mixed in $^4$He by the tensor force like a deuteron-type correlation \cite{myo09,horii12}.
In neutron-rich side, the specific 2p2h configurations above and the $p_{1/2}$ occupation of extra neutrons can couple each other, which suppresses the tensor correlation of nuclei.
This is the Pauli blocking between the 2p2h configurations and the $p_{1/2}$ extra neutrons. 
The tensor correlation depends on the configuration, which affects the splitting energy of the $p$ orbital states in the neutron-rich He and Li isotopes.
In $^{11}$Li, this tensor-type blocking naturally explains the halo formation with a large mixing of $s_{1/2}$ component \cite{myo07_11}.

In the previous study \cite{myo14}, we have applied TOSM to $^8$Be and investigated how the TOSM describes two-kinds of the structures of the shell-model and $\alpha$ clustering type states in $^8$Be.
The $^8$Be energy spectrum shows two groups. One is the ground-band three states consisting of $0^+_1$, $2^+_1$ and $4^+_1$.
These three states are recognized as the two-$\alpha$ cluster states~\cite{ikeda68,pieper04}.
The other is the excited states starting from $2^+_2$ with $E_x=16.6$ MeV experimentally. 
Above this state, there are many states with relatively small decay widths less than about 1 MeV, and their decay processes are not always $\alpha$ emission.
In the highly excited states, experimentally the $T$=1 states are located very close to the $T$=0 states.
In TOSM, we have nicely described these properties of $^8$Be including the level order of $T$=0 and $T$=1 states in the highly excited states.
One of the failures in TOSM is the small energy distance between the ground band head ($0^+_1$) and the highly excited state of $2^+_2$ ($T=0$),
which is 10.2 MeV in TOSM while 16.6 MeV in the experiment.
The relative energy between the ground band $4^+_1$ state and the $2^+_2$ state is 2 MeV which is smaller than the experimental value of 6 MeV.

The difference of the relative energy is considered to come from the insufficient description of two-$\alpha$ clustering structure in the ground band states using the shell model basis.
It was also pointed out that the tensor force contribution is stronger in the ground band states than the highly excited states, which can also be related to the $\alpha$ clustering in the ground band states.
In the highly excited states, the $T$=0 states has a rather larger tensor force contribution than the values of the $T$=1 states.
This trend can be related to the isospin dependence of the tensor force, namely one-pion exchange nature of nuclear force.
It is interesting to perform the similar analysis for other nuclei neighboring to $^8$Be focusing on the tensor correlation.

In this study, on the basis of the successful results of $^8$Be, we proceed the analysis of the states of $^{9,10}$Be and $^{10}$B using TOSM
and clarify the different structures of these nuclei as function of the excitation energy.
We investigate the structures of each states of three nuclei from the viewpoint of the tensor force in analogy with $^8$Be.
These nuclei are the systems in which one or two nucleons are added to $^8$Be and then we perform the similar analysis of $^8$Be for these nuclei.
In this study, we use the same effective interaction as the one for $^8$Be, based on the bare $NN$ interaction AV8$^\prime$.
This interaction is defined to reproduce the results of few-body calculation of $^4$He with TOSM wave function and possesses the characteristics of the bare $NN$ interaction.
We also investigate how the tensor force determines the energy spectrum of $^{9,10}$Be and $^{10}$B
in comparison with the results using the effective Minnesota interaction without the tensor force.

In Sec.\,\ref{sec:model}, we explain the TOSM.
In Sec.\,\ref{sec:result}, we give the results of $^{9,10}$Be and $^{10}$B, and discuss the role of tensor force in each energy levels of three nuclei.
A summary is provided in Sec.~\ref{sec:summary}.

\section{Method}\label{sec:model}

\subsection{tensor-optimized shell model (TOSM)}
We use the tensor-optimized shell model (TOSM) for $p$-shell nuclei.
We define a standard shell-model state for $p$-shell nuclei with mass number $A$ in order to prepare the TOSM configurations. 
The standard shell-model state $\Psi_{\rm S}$ is dominated by the low-momentum component and is given as 
\begin{eqnarray}
\Psi_{\rm S}=\sum_{k_{\rm S}} A_{k_{\rm S}}|(0s)^{4}(0p)^{A-4};k_{\rm S}\ket~.
\label{eq:standard}
\end{eqnarray}
Here, the $s$ shell is closed and the $p$ shell is open. The index $k_{\rm S}$ is the label to distinguish the configurations with the amplitude $A_{k_{\rm S}}$.
For each shell-model states, the tensor force can excite two nucleons in the $s$ and $p$ shells into various two-particle states with high-momentum components. 
In this sense, we take the configurations up to the two particle-two hole (2p2h) excitations connected with the standard shell-model configurations, which explains the concept of TOSM.

The high-momentum components brought by the strong tensor force are included in the TOSM via the excitations of two nucleons from the $s$ and $p$ shells in $\Psi_{S}$ 
to the higher shells in the 2p2h states. We have 2p2h configurations in TOSM as
\beq
\kets{{\rm 2p2h};k_{2}}=|(0s)^{n_{s}}(0p)^{n_{p}}({\rm higher})^{2};k_{2}\ket .
\eeq
We put the constraints $n_{s}+n_{p}=A-2$ and $2 \le n_{s} \le 4$.
The index $k_{2}$ is the label for various 2p2h states.
Here ``higher'' means higher shells in the the particle states, above the $s$ and $p$ shells in TOSM.

We also include the 1p1h excitations in addition to the 2p2h ones.
\beq
\kets{{\rm 1p1h};k_{1}}=|(0s)^{n_{s}}(0p)^{n_{p}}({\rm higher})^{1};k_{1}\ket
\eeq
with $n_{s}+n_{p}=A-1$ and $2 \le n_{s} \le 4$. 
These 1p1h states can bring the high-momentum components and also improve the standard shell-model state $\Psi_{\rm S}$ in the radial components.

In TOSM, in addition to the 1p1h and 2p2h states defined above,
we further extend the standard shell-model states $\Psi_{\rm S}$ in Eq. (\ref{eq:standard}) in order to allow two-particle excitations from the $s$ to $p$ shells.
This excitation is to take into account the part of the tensor correlation within the $s$ and $p$ shells.
We write these states with extension as $\kets{{\rm 0p0h};k_{0}}$ where no excitation to the particle states above the $s$ and $p$ shells.
We include these basis states in the TOSM configurations instead of $\Psi_{\rm S}$.
We explicitly define these extended shell-model states as
\beq
\kets{{\rm 0p0h};k_0}=|(0s)^{n_{s}}(0p)^{n_{p}};k_{0}\ket .
\label{eq:0p0h}
\eeq
Here, $n_{s}+n_{p}=A$ and $2 \le n_{s} \le 4$.  The index $k_{0}$ is the label for the $0p0h$ configurations. 

Superposing the 0p0h, 1p1h and 2p2h configurations, we give the total wave function $\Psi$ of TOSM as
\begin{eqnarray}
\Psi &=& \sum_{k_0} A_{k_0} \kets{{\rm 0p0h};k_0} + \sum_{k_1} A_{k_1} \kets{{\rm 1p1h};k_1} 
\nonumber\\
&+& \sum_{k_2} A_{k_2} \kets{{\rm 2p2h};k_2}~.
      \label{eq:config}
\end{eqnarray}
Three-kinds of the amplitudes $\{A_{k_0},A_{k_1},A_{k_2}\}$ are variationally determined by using the minimization of the total energy.

We describe how to construct the radial wave functions of single nucleon in each configuration of TOSM.
The 0p0h states are expressed using the harmonic oscillator wave functions as the ordinary shell model states.
The $0s$ and $0p$ basis wave functions are involved and their length parameters are determined independently and  variationally for the total energy.
The 0p0h states using the $s$ and $p$ shells become the dominant part of the wave function for $p$-shell nuclei.
For intruder states such as the positive parity states in $^9$Be, some nucleons occupy the $sd$-shell as the dominant configurations,
and the 2p2h excitations from each dominant configuration can produce the strong tensor correlation.
However, we do not include the $sd$-shell in the 0p0h configurations because the model space for $sd$-shell becomes huge.

The 1p1h and 2p2h states have the particle states above the $s$ and $p$ shells.
For particle states, we adopt the Gaussian basis functions for the description of the single-particle basis states which can have the high-momentum owing to the tensor force~\cite{hiyama03, aoyama06}.
The Gaussian basis functions are sufficient to express the high-momentum components of single nucleon by adjusting the length parameters \cite{myo09}.
This technique has been used in the previous studies of TOSM and also the cluster models \cite{myo14_2}.

We employ a sufficient number of Gaussian basis functions with various length parameters, so that the radial components of the particle states are fully described in each TOSM configuration.
The particle states are orthogonalized each other and also to the hole states of $0s$ and $0p$ shells \cite{myo07,myo09}. 
We construct the orthonormalized basis function for the particle states in terms of a linear combination of non-orthogonal Gaussian bases.
The particle wave functions can contain the high-momentum components caused by the tensor force with possible orbital angular momenta until the total energy converges.
In the numerical calculation, the partial waves of the basis states are takes up to $L_{\rm max}$.  
We take $L_{\rm max}$ as 10 to obtain the total energy convergence.
For the number of the Gaussian basis functions, typically at most 10 basis functions are used with various range parameters.

It is noted that the configuration probabilities and occupation numbers of each orbit in the state are calculated from the summation of all orbits having orthogonal radial behaviors.

We use a Hamiltonian with a bare $NN$ interaction for a mass number $A$
\begin{eqnarray}
    H
&=& \sum_i^{A} T_i - T_{\rm c.m.} + \sum_{i<j}^{A} V_{ij} , 
    \label{eq:Ham}
    \\
    V_{ij}
&=& v_{ij}^C + v_{ij}^{T} + v_{ij}^{LS} + v_{ij}^{Clmb} .
\end{eqnarray}
Here, $T_i$ and $T_{\rm c.m.}$  are the kinetic energies of each nucleon and a center-of-mass part.
We use a bare interaction $V_{ij}$ such as AV8$^\prime$ \cite{pudliner97} consisting of central $v^C_{ij}$, tensor $v^T_{ij}$, and spin-orbit $v^{LS}_{ij}$ terms, and $v_{ij}^{Clmb}$ the Coulomb term.  

We treat the excitations of the center-of-mass considering the Hamiltonian of the center-of-mass motion in the Lawson method~\cite{lawson}. 
In this study, we use the value of $\hbar \omega$ for the center-of-mass motion as the averaged one of the $0s$ and $0p$ orbits in the 0p0h states considering 
the weight of the occupation numbers in each orbit \cite{myo11}.
Adding this center-of-mass Hamiltonian to the original Hamiltonian in Eq.~(\ref{eq:Ham}), we can effectively exclude the excitation of the center-of-mass motion.  

The energy variation for the TOSM wave function $\Psi$ defined in Eq.~(\ref{eq:config}) is performed 
with respect to the two kinds of variational parameters; one is the length parameters of the Gaussian basis functions
and the other is the amplitudes of the TOSM configurations, $A_{k_0}$, $A_{k_2}$ and $A_{k_2}$ given in Eq.~(\ref{eq:config}).

We employ UCOM to take into account the short-range correlation coming from the $NN$ interaction \cite{feldmeier98,roth10}. 
We define the wave function $\Phi$ including the short-range correlation by using the TOSM wave function $\Psi$ as $\Phi=C\Psi$, where the unitary operator $C$ is defined as
\begin{eqnarray}
C     &=&\exp(-i\sum_{i<j} g_{ij}) .
\label{eq:ucom}
\end{eqnarray}
We take the Hamiltonian transformed as $C^\dagger H C$ using UCOM within the two-body level \cite{feldmeier98} considering the nature of short-range correlation.

The two-body operator $g$ in Eq.~(\ref{eq:ucom}) is given as
\begin{eqnarray}
g &=& \frac12 \left\{ p_r s(r)+s(r)p_r\right\} ,
\label{eq:ucom_g}
\end{eqnarray}
where $p_r$ is the operator for the radial part of the relative momentum. 
The function $s(r)$ is determined to minimize the total energy of nuclei.
In the present analysis, we commonly use the same $s(r)$ functions determined in $^4$He \cite{myo11,myo12}.
To simplify the calculation, we adopt the ordinary UCOM instead of the $S$-UCOM.
The $S$-UCOM extensively introduces the partial-wave dependence of $s(r)$, such as two-kinds of $s(r)$'s for the $s$-wave and for other partial waves \cite{myo09}.
This $S$-UCOM can improve the behavior of the relative $d$-wave states of nucleon pair in the short-range part.
As a result, the $sd$ coupling and the total energy are increased in $S$-UCOM than the UCOM case.

\subsection{$NN$ interaction}

In this paper, we use two-kinds of $NN$ interactions for comparison; one is the bare AV8$^\prime$ interaction with central, $LS$ and tensor forces.
This interaction is used for the benchmark calculation of $^4$He given by Kamada et al. without the Coulomb force \cite{kamada01}.  
The other is the effective Minnesota (MN) $NN$ interaction, which does not have the tensor force. 
For MN interaction, we choose the $u$ parameter as 0.95 for the central part and use set III of the $LS$ part \cite{tang78}.
In case of MN interaction with Coulomb interaction and without $LS$ force, binding energy of $^4$He with TOSM is 29.72 MeV, which is very close to the rigorous calculation of 29.94 MeV \cite{varga95}.

\begin{table}[t]
\centering
\caption{Hamiltonian components in TOSM, TOFM and SVM for $^4$He in MeV. In the last row of TOSM, AV8$^\prime_{\rm eff}$ is used \cite{myo14}.}
\begin{tabular}{l|ccccc}
\noalign{\hrule height 0.5pt}
                                &  Energy  &  Kinetic &  Central & Tensor    &  $LS$   \\
\noalign{\hrule height 0.5pt}
TOSM \cite{myo09}               & $-22.30$ & $ 90.50$ & $-55.71$ & $-54.55$  & $-2.53$ \\
SVM~\cite{kamada01}             & $-25.92$ & $102.35$ & $-55.23$ & $-68.32$  & $-4.71$ \\ 
TOFM \cite{horii12}             & $-24.18$ & $~95.50$ & $-54.67$ & $-61.32$  & $-4.09$ \\   
TOSM with AV8$^\prime_{\rm eff}$& $-26.16$ & $~95.45$ & $-56.17$ & $-62.43$  & $-3.02$ \\
\noalign{\hrule height 0.5pt}
\end{tabular}
\label{tab:4He_ham}
\end{table}

In TOSM with AV8$^\prime$ interaction as shown in Table \ref{tab:4He_ham}, it is found that the tensor and $LS$ contributions for $^4$He give smaller values than those in the stochastic variational method (SVM) using correlated Gaussian basis functions \cite{varga95}, which is one of the rigorous calculations~\cite{kamada01}.
One of the reasons for the shortage in TOSM is the contributions from higher configurations beyond the 2p2h ones.
The other reason is the two-body truncation of the transformed Hamiltonian with UCOM.
There remains a small contribution in the short-range part of the tensor force and also of the $LS$ force, 
which can couple with the short-range UCOM and produces the many-body UCOM term\cite{feldmeier98}.

Horii et al. examined the energy of this coupling with the few-body SVM \cite{horii12} without UCOM.
They also propose the one-pair type of the coupling induced by the tensor force introducing the single $Y_2$ function in the global vector for angular momentum,
which produces the $d$-wave component in the wave function \cite{horii12}. This model is called the tensor-optimized few-body model (TOFM).
The physical concept between TOFM and TOSM is the same except for the short-range UCOM part in TOSM.
In TOFM, the short-range correlation is directly treated in the wave function. On the other hand, this correlation is approximately taken into account in the TOSM using UCOM.
It is shown that TOFM gives a good binding energy of $^4$He with AV8$^\prime$ as shown in Table \ref{tab:4He_ham} as compared with the benchmark calculation (SVM).
The energy of TOFM is lower than the value of TOSM by about 2 MeV \cite{horii12} and this difference comes from the use of UCOM in TOSM.
From this result, the three-body UCOM term is considered to recover the missing energy of TOSM~\cite{myo09, feldmeier98}. 

In the previous analysis of $^8$Be \cite{myo14}, we have considered the shortages of the tensor and $LS$ contributions in TOSM. 
We introduced the effective interaction based on the AV8$^\prime$, which keeps the characteristics of the $NN$ interaction as much as possible.
We include these effects in TOSM to reproduce the TOFM results of $^4$He as closely as possible by adjusting the tensor and $LS$ matrix elements phenomenologically. 
We enhance the tensor matrix elements by 10\% with the enhancement factor $X_{\rm T}$=1.1. For the $LS$ matrix elements, we use $X_{LS}$=1.4 for 40\% enhancement.
Using this interaction, the total energy is nicely obtained and more than 90\% of the tensor and kinetic energy components of the SVM calculation is reproduced as shown in Table \ref{tab:4He_ham}.
In addition, the TOSM solutions almost simulate those of TOFM.
These results indicate that the missing components of TOSM are recovered by using the effective interactions.
In the present analysis, we use this interaction called as ``AV8$^\prime_{\rm eff}$''.
This AV8$^\prime_{\rm eff}$ interaction successfully describes the level order of $^8$Be \cite{myo14}, 
but it is not obvious whether this interaction is applicable to the systematic description of the $p$-shell nuclei.
Hence, we examine the applicability of this interaction using the present analysis of the $p$-shell nuclei with TOSM.

\section{Results}\label{sec:result}

\subsection{$^9$Be}\label{sec:Be9}

\begin{figure}[b]
\centering
\includegraphics[width=7.0cm,clip]{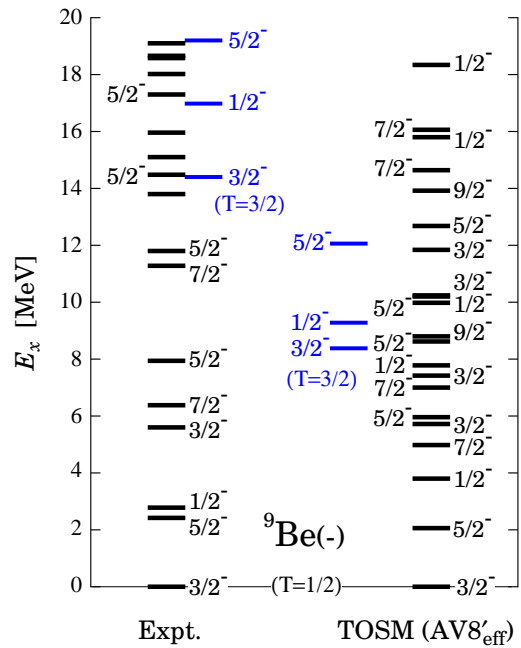}
\caption{Excitation energy spectrum of $^9$Be using AV8$^\prime_{\rm eff}$.
Left-hand side are experimental data for $T$=1/2 (black line) and $T$=3/2 (blue lines).  
Right-hand side are results of TOSM.}
\label{fig:AV_Be9}
\end{figure}

\begin{figure}[t]
\centering
\includegraphics[width=5.5cm,clip]{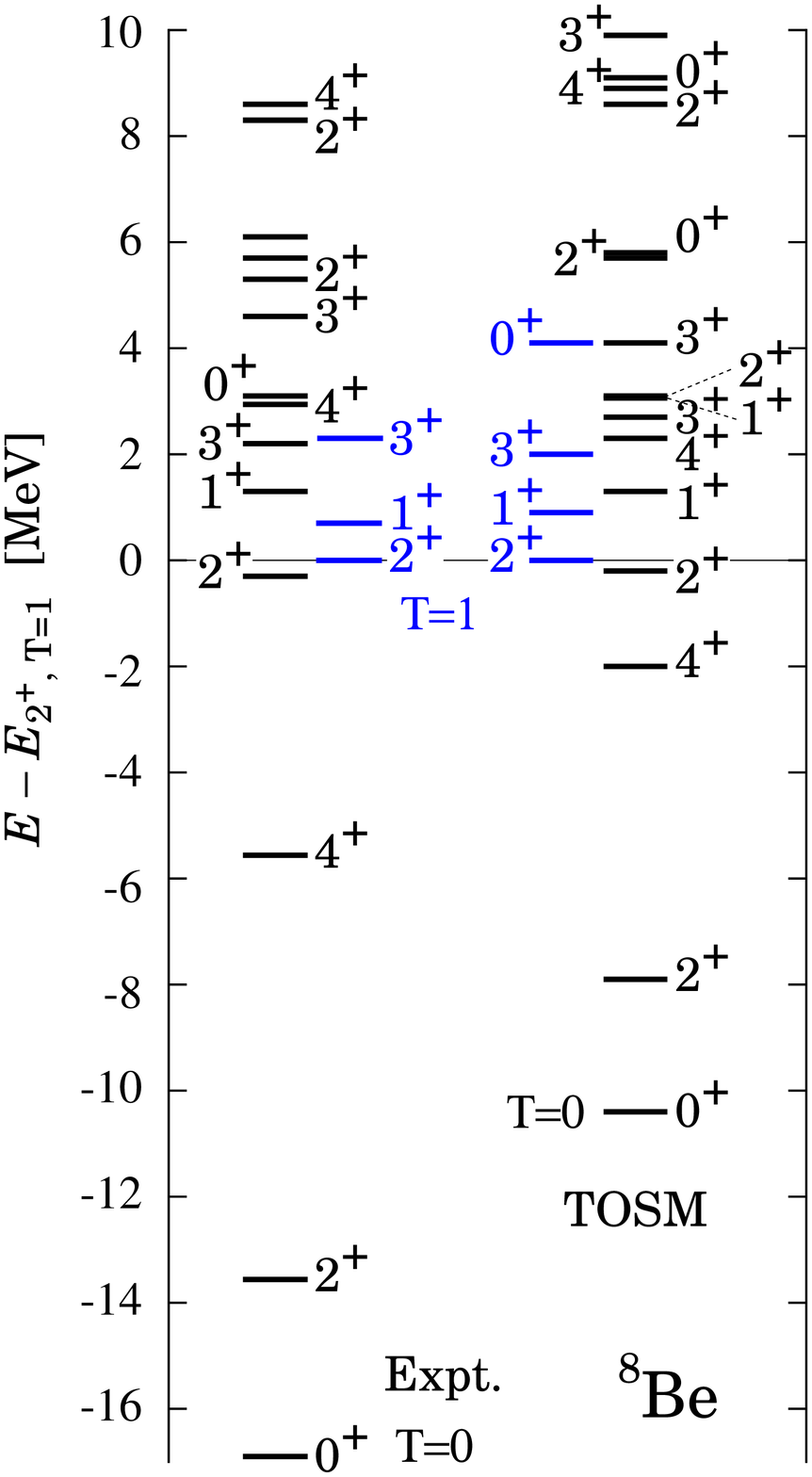}
\caption{Excitation energy spectrum of $^8$Be using AV8$^\prime_{\rm eff}$ normalized to the $2^+$($T$=1) state.}
\label{fig:AV_Be8}
\end{figure}

\begin{figure}[t]
\centering
\includegraphics[width=7.0cm,clip]{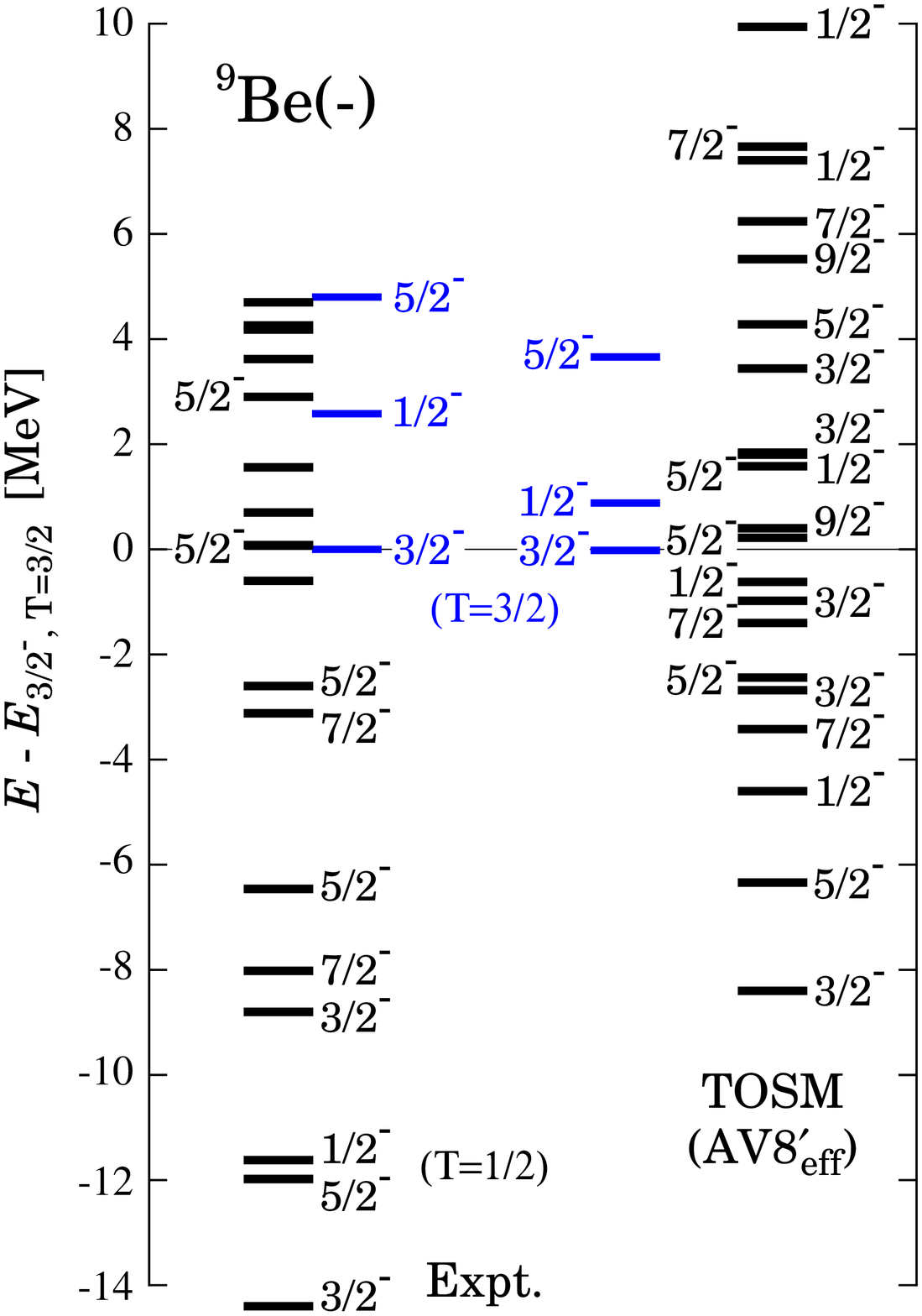}
\caption{Excitation energy spectrum of $^9$Be using AV8$^\prime_{\rm eff}$ normalized to the $\frac32^-$($T$=3/2) state.}
\label{fig:AV_Be9_2}
\end{figure}

We discuss the structures of $^9$Be with negative parity states in TOSM using AV8$^\prime_{\rm eff}$, which is used in the analysis of $^8$Be. 
The total binding energy of the ground state of $^9$Be is obtained as $27.71$ MeV in TOSM, which is largely smaller than the experimental value $58.17$ MeV.
This difference mainly comes from the missing higher configurations beyond the 2p2h excitations in TOSM and also the three-nucleon interaction.
The $\alpha$ clustering correlation in $^9$Be is also considered to explain the energy difference \cite{myo14}.

We show the excitation energy spectra of $^9$Be in Fig. \ref{fig:AV_Be9} for $T$=1/2 and 3/2.
In the experimental spectrum, the spins of most of the highly excited states are not confirmed yet for $T=1/2$.
It is found that there can be two groups of states in $^9$Be for the experimental spectrum; one is up to the excitation energy of 12 MeV and 
the other is the highly excited states starting from 14 MeV, which are degenerated with the $T$=3/2 states.
The relative energy between $5/2^-_3$ and the state with $E_x=13.8$ MeV in $^9$Be is about 2 MeV in the experiments, while these state are overlapped in some region in the calculated spectrum.
In the TOSM results, for $T$=1/2, we almost reproduce the experimental level order of the low-lying states starting from the ground $\frac32^-$ states.
For the $T$=3/2 states the excitation energies are lower than the experimental values by about 6 MeV.

Here we discuss the analogy of level structures between $^9$Be and $^8$Be.
The binding energy of $^8$Be with AV8$^\prime_{\rm eff}$ is obtained as $30.19$ MeV, which is smaller than the experimental value $56.50$ MeV.
In the previous analysis of $^8$Be \cite{myo14}, similarly we have obtained the two groups in the spectrum; one is the three ground band states of $0^+$, $2^+$ and $4^+$ 
and the other is the highly excited states starting from the $2^+_2$ state at 16.6 MeV of the excitation energy as shown in Fig. \ref{fig:AV_Be8}.
In the figure the spectrum is normalized at the $2^+$ ($T=1$) state of $^8$Be, which is regarded as the shell-like state.
In TOSM, the relative energy between $4^+_1$ and $2^+_2$ is about 2 MeV and smaller than the experimental value of 6 MeV.
The main reason of difference is considered to be the missing energy coming from the $\alpha$ clustering component in the ground band states of $^8$Be.
We consider the same effect on the $^9$Be spectrum. 
For this purpose, we renormalize the spectrum to the $T$=3/2, $J^\pi$=$\frac32^-$ state as shown in Fig. \ref{fig:AV_Be9_2}.
It is found that there is a good correspondence in the $T$=3/2 states between TOSM and the experiments.
This indicates that the TOSM nicely described the $T$=3/2 states and the states are considered to be mainly the shell-like states, as was obtained in the $^9$Li \cite{myo12}.

For $T$=1/2 states, it is found that the energy of the ground state is missed by about 6 MeV in comparison with the experiments in Fig. \ref{fig:AV_Be9_2}.
This value is close to the $^8$Be case as shown in Fig. \ref{fig:AV_Be8}.
In the $^8$Be case, we have discussed the $\alpha$ clustering effect using two-$\alpha$ cluster model and the possible energy gain from the $\alpha$ clustering is estimated as 5 MeV \cite{myo14}.
For $^9$Be case, it is expected that the inclusion of the $\alpha$+$\alpha$+$n$ three-body component may help the lack of the energy of $^9$Be in TOSM for low-lying states.
These results imply that the mixture of the $\alpha$ cluster component in the TOSM basis states is desirable to improve the energy spacing of two groups in $^9$Be.
It is interesting to develop the TOSM to include the $\alpha$ clustering correlation explicitly and to express the tensor contribution in each $\alpha$ particle. 

\begin{table}[t]
\centering
\caption{RMS radii of $^{9,10}$Be and $^{10}$B in the units of fm.}
\begin{tabular}{c|ccccc}
\noalign{\hrule height 0.5pt}
               &  TOSM    & Experiment\cite{tanihata88} \\
\noalign{\hrule height 0.5pt}
$^8$Be         &  2.21    & ---     \\
$^9$Be         &  2.32    & 2.38(1) \\
$^{10}$Be      &  2.31    & 2.30(2) \\
$^{10}$B       &  2.20    & ---     \\
\noalign{\hrule height 0.5pt}
\end{tabular}
\label{tab:radius}
\end{table}

For the ground state of $^9$Be, the matter radius is obtained as 2.32 fm in TOSM as shown in Table \ref{tab:radius},
which is slightly smaller than the experimental value of 2.38(1) fm \cite{tanihata88}. This trend of radius occurs in the case of $^8$Be (2.21 fm) \cite{myo14} in comparison with the two $\alpha$ cluster model (2.48 fm).
The small radii of $^8$Be and $^9$Be in TOSM indicate that the $\alpha$ clustering correlation is not fully included in the present solution of TOSM.
In the shell model, it is generally difficult to express the asymptotic form of the spatially developed $\alpha$ clustering states.
Naively, many particle-many hole excitations in the shell model bases might be necessary to assist the formation of the well-separated two-$\alpha$ clusters in space.
In TOSM, the 2p2h excitation is used to incorporate the tensor correlation in the single $\alpha$ particle with high-momentum components by about 10\% \cite{myo09,myo11}.
This indicates that when two $\alpha$ particles are established in $^8$Be and $^9$Be, each $\alpha$ particle independently needs the 2p2h components to express the tensor correlation,
although the probability of this situation is expected not so high.
In the TOSM, the approximation of the 2p2h excitations might restrict the spatial cluster formation in $^8$Be and $^9$Be.

The TOSM with AV8$'_{\rm eff}$ is found to reproduce entirely well the energy level order of $^9$Be.
We examine how the tensor matrix elements contribute to determine the level order of $^9$Be.
We investigate the level structures of $^9$Be by changing the strengths of tensor forces from the value of AV8$^\prime_{\rm eff}$.
The same analysis was performed for $^8$Be \cite{myo14}.

\begin{figure*}[t]
\centering
\includegraphics[width=12.0cm,clip]{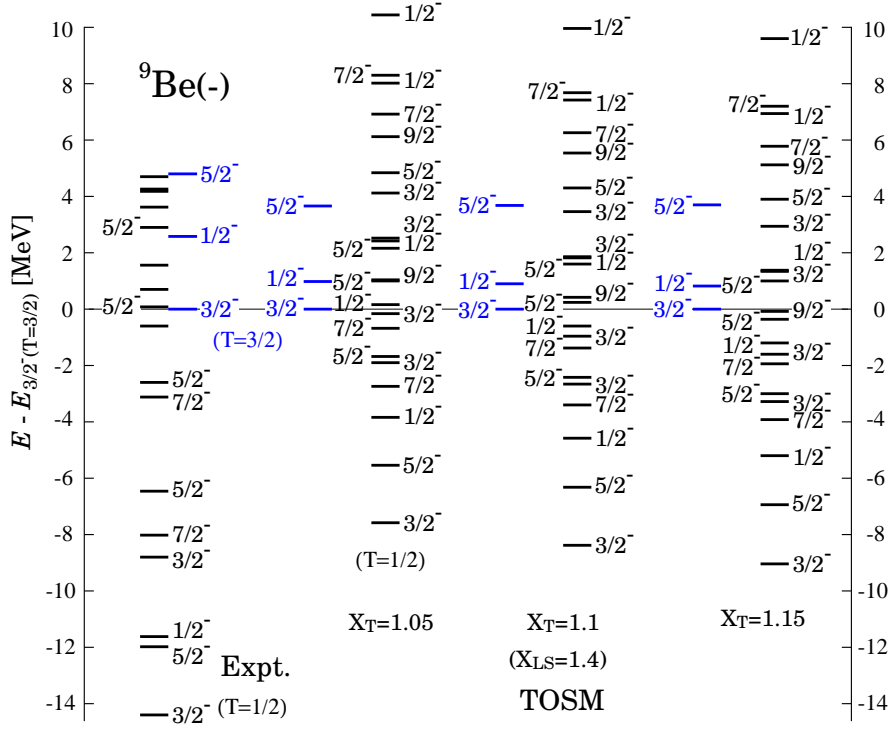}
\caption{Energy spectrum of $^9$Be in TOSM by changing the tensor strength about $\pm 5\%$ from AV8$^\prime_{\rm eff}$.
The black lines denote the $T$=1/2 states and the blue lines the $T$=3/2 states.} 
\label{fig:AV_Be9_tensor}
\end{figure*}

We begin with the effective interaction AV8$^\prime_{\rm eff}$, namely, $X_{\rm T}$=1.1 and change the tensor strength.
The result is shown in Fig. \ref{fig:AV_Be9_tensor}, normalized to the $\frac32^-$ ($T$=3/2) state.
It is confirmed that the energy spacings do not change so much in this parameter range of $X_{\rm T}$.
It is found that the relative energy between the low-lying $T$=1/2 states and the three $T$=3/2 states increases as the tensor force is strengthened.
This result indicates that the $T$=1/2 states can be easily affected by tensor force than the $T$=3/2 states.
This is related to the attractive nature of the $T$=0 channel of the tensor force. 
Hence, as the tensor correlation becomes stronger, the smaller isospin $T$=1/2 states gain more energy than the larger isospin $T$=3/2 states.
In addition, the level orders in each isospin state do not change so much. 
This is considered as follows. The tensor force strongly couples the 0p0h and 2p2h configurations, in which the particle states involve high-momentum components. 
This coupling induced by the tensor force contributes to the total energy of every state commonly within the same isospin state.  
Hence, the low-lying relative spectra of $^9$Be does not show the strong dependence of the tensor force strength.

The Hamiltonian components of each state are discussed later.
The same effect of tensor force on the isospin of nuclei is confirmed in $^8$Be \cite{myo14}, in which the $T$=0 and $T$=1 states are compared
and the $T=0$ states tends to gain the energy more as the tensor correlation becomes stronger.
From this result of $^9$Be, it is found that the tensor force affects the relative energies between the $T$=1/2 and $T$=3/2 states.
This is just the same conclusion obtained for $^8$Be \cite{myo14}. 

\begin{figure}[t]
\centering
\includegraphics[width=7.0cm,clip]{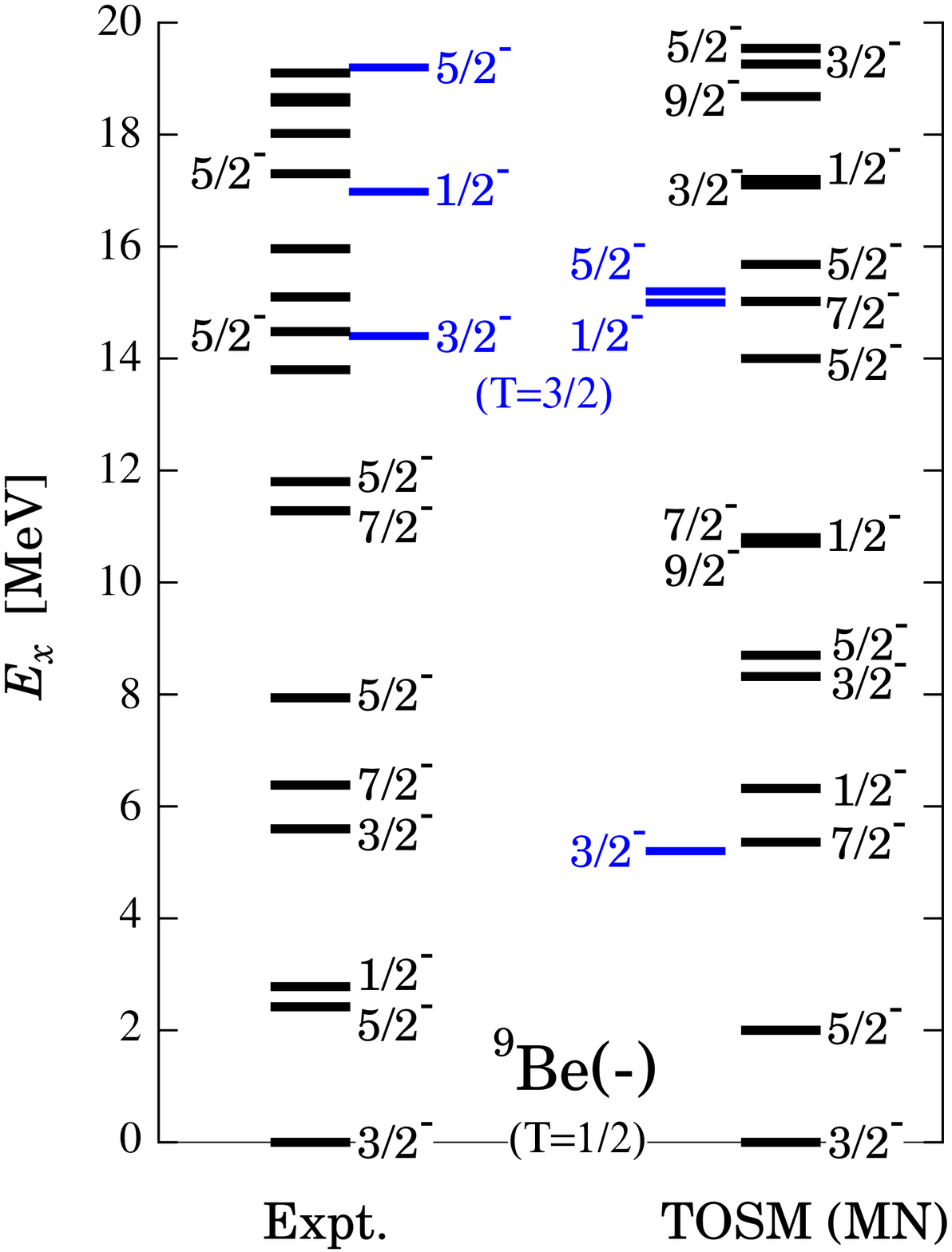}
\caption{Excitation energy spectrum of $^9$Be in TOSM using the Minnesota interaction (MN).}
\label{fig:MN_Be9}
\end{figure}

We also show the energy spectrum of $^9$Be using the effective MN interaction to see the effect of the tensor force on the spectrum.
In this study, we reduce the strength of the $LS$ force in the MN interaction by 30\% to give the same $LS$ splitting energy 1.5 MeV in $^5$He using AV8$^\prime$ in TOSM \cite{myo11,myo14}.
The binding energy of the $^9$Be ground state is obtained as $66.50$ MeV, which is close to the experimental value of $58.17$ MeV.
This interaction gives a radius of 1.96 fm for $^9$Be much smaller than the AV8$^\prime_{\rm eff}$ case of 2.32 fm .
Hence, the saturation property cannot be reproduced in the MN interaction.

We show the excitation energy spectrum with MN in Fig.\,\ref{fig:MN_Be9}.
The spectrum reproduces the overall trend of the experiments including the level density.
However, the level order is different from the experiment in some part.
At lower excitation energies of the $T$=1/2 states , it is found that the $\frac12^-_1$ is highly located than the experimental value.
For the $T$=3/2 states, the location of the $\frac32^-$ state is much different from the experiment while other two states of $\frac12^-$ and $\frac52^-$ are close to the experiments. 
From these results obtained using MN interaction, there are differences in the excitation energies of the some specific states.
As compared with the MN results, the AV8$^\prime_{\rm eff}$ interaction gives the reasonable energy spectrum of $^9$Be for two isospin states.

\begin{table}[th]
\centering
\caption{Hamiltonian components of $^9$Be for $T$=1/2 and 3/2 states in MeV.}
\begin{tabular}{c|ccccc}  
\noalign{\hrule height 0.5pt}
State($T$=1/2) &  Energy  &  Kinetic    &  Central & Tensor    &  $LS$   \\
\noalign{\hrule height 0.5pt}
$\frac12^-_1$ & $-23.92$ & $ 214.75$ & $-126.17$ & $-101.09$ & $-11.40$   \\
$\frac12^-_2$ & $-19.92$ & $ 212.31$ & $-114.46$ & $-101.06$ & $-16.72$   \\
$\frac12^-_3$ & $-17.51$ & $ 210.56$ & $-113.92$ & $-97.77 $ & $-16.38$   \\
\noalign{\hrule height 0.5pt}
$\frac32^-_1$ & $-27.71$ & $ 217.49$ & $-127.10$ & $-102.07$ & $-16.03$  \\
$\frac32^-_2$ & $-21.98$ & $ 213.46$ & $-123.70$ & $-101.25$ & $-10.47$  \\
$\frac32^-_3$ & $-20.28$ & $ 211.81$ & $-115.56$ & $-101.83$ & $-14.69$  \\
\noalign{\hrule height 0.5pt}
$\frac52^-_1$ & $-25.64$ & $ 216.57$ & $-124.50$ & $-101.49$ & $-16.23$  \\
$\frac52^-_2$ & $-21.74$ & $ 213.69$ & $-122.08$ & $-100.36$ & $-12.99$  \\
$\frac52^-_3$ & $-19.07$ & $ 211.14$ & $-113.16$ & $-100.95$ & $-16.10$  \\
\noalign{\hrule height 0.5pt}
$\frac72^-_1$ & $-22.73$ & $ 214.93$ & $-121.75$ & $-99.08$  & $-16.82$  \\
$\frac72^-_2$ & $-20.71$ & $ 212.86$ & $-114.29$ & $-99.93$  & $-19.37$  \\
$\frac72^-_3$ & $-16.72$ & $ 210.57$ & $-115.92$ & $-98.45$  & $-12.92$  \\
\noalign{\hrule height 0.5pt}
$\frac92^-_1$ & $-18.91$ & $212.67$ & $-117.4589$ & $-98.31$  &  $-15.81$  \\    
$\frac92^-_2$ & $-13.78$ & $208.87$ & $-113.1296$ & $-96.90$  &  $-12.62$  \\
\noalign{\hrule height 0.5pt}
\end{tabular}
\vspace*{0.1cm}
\\
\begin{tabular}{c|ccccc}   
\noalign{\hrule height 0.5pt}
State($T$=3/2)&  Energy  &  Kinetic  &  Central  & Tensor    &  $LS$   \\
\noalign{\hrule height 0.5pt}
$\frac12^-$       & $-17.59$  &  $187.78$  & $-106.57$ &  $-85.95$ & $-12.85$ \\
$\frac32^-$       & $-18.50$  &  $189.10$  & $-107.86$ &  $-84.25$ & $-15.49$ \\
$\frac52^-$       & $-14.81$  &  $187.48$  & $-106.51$ &  $-83.83$ & $-11.94$ \\
\noalign{\hrule height 0.5pt}
\end{tabular}
\label{tab:9Be_ham}
\end{table}

We discuss the Hamiltonian components in each state of $^{9}$Be using AV8$^\prime_{\rm eff}$ in TOSM to understand the contributions of the tensor force explicitly.
In Table \ref{tab:9Be_ham}, we show the Hamiltonian components of the $^9$Be states.

We focus on the differences between the $T$=1/2 and $T$=3/2 states of $^9$Be.
In Table \ref{tab:9Be_ham}, it is shown that the $T$=3/2 states show the smaller tensor contributions than the $T$=1/2 case by about 15 MeV.
The kinetic energies also have a similar trend owing to the high-momentum component brought by the tensor force.
This result can have the relation to the isospin dependence of the tensor force, 
because the $T$=1/2 states can easily contain the $T$=0 nucleon pair states which induce the stronger tensor correlation than that of the $T$=1 nucleon pair.
This results is consistent to the dependence of the energy spectrum on the tensor strength as shown in Fig.~\ref{fig:AV_Be9_tensor}.
We have confirmed the same feature of the isospin dependence of the tensor and kinetic contributions in $^8$Be \cite{myo14}.

It is found that the $\frac32^-_1$ ground state with $T$=1/2 possesses the largest tensor contribution of $-102$ MeV and also the largest kinetic energy and central contributions.
As the excitation energy goes up, the states tend to reduce the matrix elements of each Hamiltonian components.
For the $LS$ contribution, this value mainly comes from the single-particle configurations, such as the $p_{1/2}$ and $p_{3/2}$ orbits.

Dominant configurations of several states in $^9$Be are shown in Tables \ref{conf9_1} for $T=1/2$ and in Tables \ref{conf9_2} for $T=3/2$.
It is found that the $T=3/2$ states have rather single configuration properties, 
and the configuration mixing occurs more strongly in the $T=1/2$ states than the $T=3/2$ states. 

\begin{table}[t]
\centering
\caption{Dominant configurations of $^9$Be in $T=1/2$ with their squared amplitudes $(A^J_k)^2$ using AV8$^\prime_{\rm eff}$ interaction.
The configuration indices $(JT)$ are the spin-isospin quantum numbers.}
\label{conf9_1}
\begin{tabular}{ll|lrc}
\noalign{\hrule height 0.5pt}
\multicolumn{2}{c|}{$\frac12^-_1$ }         & \multicolumn{2}{c}{$\frac12^-_2$ }                             \\ \hline
$(0s)^4(0p_{3/2})^4_{00}(0p_{1/2})$  & 0.49 & $(0s)^4(0p_{3/2})^4_{11}(0p_{1/2})$                     & 0.36 \\
$(0s)^4(0p_{3/2})^2_{01}(0p_{1/2})^3$& 0.09 & $(0s)^4(0p_{3/2})^5$                                    & 0.25 \\
                                     &      & $(0s)^4(0p_{3/2})^3_{\frac32,\frac12}(0p_{1/2})^2_{10}$ & 0.10 \\
\noalign{\hrule height 0.5pt}
\end{tabular}
\vspace*{0.1cm}
\\
\begin{tabular}{ll|lrc}
\noalign{\hrule height 0.5pt}
\multicolumn{2}{c|}{$\frac32^-_1$ }                           & \multicolumn{2}{c}{$\frac32^-_2 $ }                            \\ \hline
$(0s)^4(0p_{3/2})^5$                                   & 0.36 & $(0s)^4(0p_{3/2})^4_{21}(0p_{1/2})$                     & 0.30 \\
$(0s)^4(0p_{3/2})^3_{\frac32,\frac32}(0p_{1/2})^2_{01}$& 0.17 & $(0s)^4(0p_{3/2})^4_{\frac32,\frac12}(0p_{1/2})^2_{01}$ & 0.29 \\
$(0s)^4(0p_{3/2})^3_{\frac12,\frac12}(0p_{1/2})$       & 0.10 &                                                         &      \\
\noalign{\hrule height 0.5pt}
\end{tabular}
\vspace*{0.1cm}
\\
\begin{tabular}{ll|lrc}
\noalign{\hrule height 0.5pt}
\multicolumn{2}{c|}{$\frac52^-_1$ }                   & \multicolumn{2}{c}{$\frac52^-_2$ }                             \\ \hline
$(0s)^4(0p_{3/2})^4_{21}(0p_{1/2})$            & 0.30 & $(0s)^4(0p_{3/2})^4_{20}(0p_{1/2})$                     & 0.37 \\
$(0s)^4(0p_{3/2})^5$                           & 0.23 & $(0s)^4(0p_{3/2})^3_{\frac32,\frac12}(0p_{1/2})^2_{10}$ & 0.16 \\
\noalign{\hrule height 0.5pt}
\end{tabular}
\vspace*{0.1cm}
\\
\begin{tabular}{ll|lrc}
\noalign{\hrule height 0.5pt}
\multicolumn{2}{c|}{$\frac72^-_1$ }                   & \multicolumn{2}{c}{$\frac72^-_2$ }                     \\ \hline
$(0s)^4(0p_{3/2})^4_{31}(0p_{1/2})$            & 0.37 & $(0s)^4(0p_{3/2})^5$                            & 0.36 \\
$(0s)^4(0p_{3/2})^5$                           & 0.26 & $(0s)^4(0p_{3/2})^4_{31}(0p_{1/2})$             & 0.31 \\
\noalign{\hrule height 0.5pt}
\end{tabular}
\end{table}

\begin{table}[t]
\centering
\caption{Dominant configurations of $^9$Be in $T=3/2$ with their squared amplitudes $(A^J_k)^2$ using AV8$^\prime_{\rm eff}$ interaction.}
\label{conf9_2}
\begin{tabular}{ll|lrc}
\noalign{\hrule height 0.5pt}
\multicolumn{2}{c|}{$\frac12^-$}                      & \multicolumn{2}{c}{$\frac32^-$}                         \\ \hline
$(0s)^4(0p_{3/2})^4_{02}(0p_{1/2})$            & 0.82 & $(0s)^4(0p_{3/2})^5$                            & 0.53  \\
$(0s)^4(0p_{3/2})^2_{02}(0p_{1/2})^3$          & 0.06 & $(0s)^4(0p_{3/2})^3_{3/2,1/2}(0p_{1/2})^2_{01}$ & 0.11  \\
\noalign{\hrule height 0.5pt}
\end{tabular}
\vspace*{0.1cm}
\\
\begin{tabular}{ll|lrc}
\noalign{\hrule height 0.5pt}
\multicolumn{2}{c|}{$\frac52^-$}                       & \multicolumn{2}{c}{$\frac72^-$}                \\ \hline
$(0s)^4(0p_{3/2})^4_{21}(0p_{1/2})$             & 0.59 & $(0s)^4(0p_{3/2})^4_{31}(0p_{1/2})$    & 0.77  \\
$(0s)^4(0p_{3/2})^3_{5/2,1/2}(0p_{1/2})^2_{01}$ & 0.10 &  &   \\
\noalign{\hrule height 0.5pt}
\end{tabular}
\end{table}

For reference, the $\frac12^+$ state of $^9$Be is obtained at the excitation energy of 8.4 MeV in TOSM, which is located higher than the experimental value by 6.7 MeV. 
The $\frac12^+$ state is observed very close to the $\alpha$+$\alpha$+$n$ threshold energy.
For this intruder state, last neutron can occupy the $1s$ orbit coupled with the $^8$Be ($0^+$) state weakly. 
In the present TOSM, the description of the dominant configuration for the intruder states is insufficient as explained in Sec.\,\ref{sec:model}, 
because the $1s$ shell is treated as the higher shells.
In the dominant 0p0h configurations for $^9$Be($\frac12^+$), one nucleon is excited from $0s$ shell to $0p$ shell in TOSM. 
The result of TOSM indicates that the inclusion of the $sd$-shell in the 0p0h configurations in Eq.~(\ref{eq:0p0h}) could improve the description of $^9$Be($\frac12^+$).

\subsection{$^{10}$Be}\label{sec:Be10}

\begin{figure}[t]
\centering
\hspace*{-0.5cm}
\includegraphics[width=8.5cm,clip]{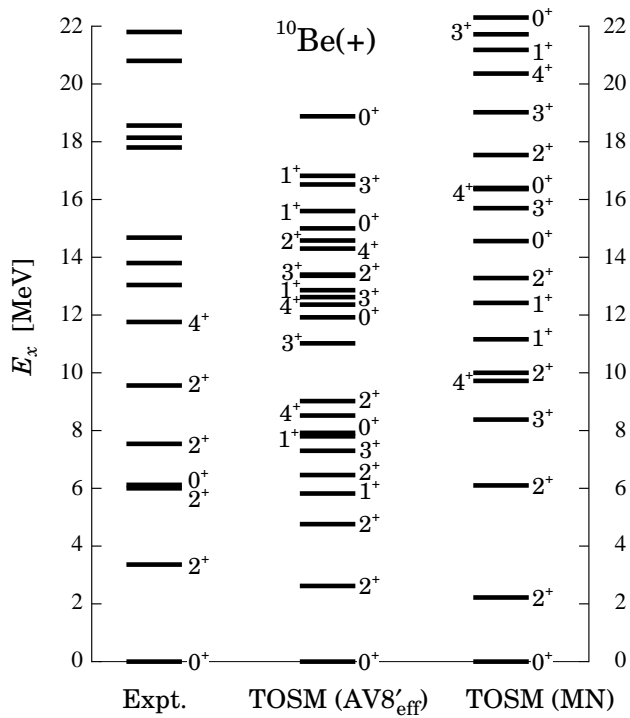}
\caption{Excitation energy spectrum of $^{10}$Be with $T=1$ using AV8$^\prime_{\rm eff}$ and MN interactions.
Left-hand side are experimental data.}
\label{fig:Be10_AV_MN}
\end{figure}

We discuss the structures of $^{10}$Be for their positive parity states with $T=1$ in TOSM.
The binding energy of $^{10}$Be  with AV8$^\prime_{\rm eff}$ is obtained as $29.91$ MeV, which is smaller than the experimental value $64.98$ MeV.
The energy spectrum of $^{10}$Be is shown in Fig. \ref{fig:Be10_AV_MN}.
Experimentally, the spins of highly excited states are not assigned yet, similar situation to the $^9$Be case.

From the comparison between TOSM and experiments, in the low-lying states,
the $0^+_2$ state is located at about 8 MeV of the excitation energy, which is 2 MeV different from the experiment.
We will focus on the structure of the $0^+$ states later.
The excitation energies of the $2^+$ states are well reproduced in TOSM from $2^+_1$ to $2^+_4$.
The $4^+_1$ state, which is considered to form the band structure with the ground and $2^+_1$ states, is located at lower excitation energy than the experiment.
This trend of the small band energy can also be seen in the case of $^8$Be in TOSM in Fig.~\ref{fig:AV_Be8} \cite{myo14}.
We also predict low-lying $1^+$ state at the excitation energy of about 5.8 MeV, which is not confirmed experimentally.
The matter radius of the ground state of $^{10}$Be is 2.31 fm in TOSM as shown in Table \ref{tab:radius}, which is close to the experimental value of 2.30(2) fm \cite{tanihata88}.

For the $0^+$ series, several theories suggest the two-$\alpha$ clustering state in the $0^+_2$ with a large mixing of the $sd$ shell of valence two neutrons
\cite{dote97,enyo99,itagaki00,arai04,ito08}.
On the other hand, in TOSM, the mixings of $sd$-shell are not large in all $0^+$ states as shown in Table \ref{tab:occ_Be10}.
This is related to the large excitation energy of $\frac12^+$ state in $^9$Be as was explained.
The $sd$-shell corresponds to the intruder orbit and the lowering of the $sd$-shell in the low-excitation energy can often be seen in the neutron-rich $p$-shell nuclei,
such as in $^{11}$Be and $^{12}$Be \cite{enyo02,ito14}.
In TOSM, it is interesting to examine this phenomenon by extending the standard shell-model state to include the $sd$-shell in Eq.~(\ref{eq:standard}), 
although this extension needs the huge model space and computational effort at present.

\begin{table}[t]
\centering
\caption{Nucleon occupation numbers of $^{10}$Be ($0^+$).}
\begin{tabular}{c|cccccccccc}
\noalign{\hrule height 0.5pt}
        &~$0s_{1/2}$&$0p_{1/2}$&$0p_{3/2}$&$1s_{1/2}$&$d_{3/2}$&$d_{5/2}$&$1p_{1/2}$&$1p_{3/2}$\\ 
\noalign{\hrule height 0.5pt}                                                 
$0^+_1$ &~3.77      & 0.92     &  4.90    & 0.04     & 0.06    & 0.05    & 0.03     & 0.05  \\ 
$0^+_2$ &~3.78      & 1.71     &  4.09    & 0.04     & 0.06    & 0.05    & 0.03     & 0.05  \\ 
$0^+_3$ &~3.78      & 1.83     &  3.97    & 0.04     & 0.06    & 0.05    & 0.03     & 0.04  \\ 
$0^+_4$ &~3.78      & 1.40     &  4.41    & 0.05     & 0.06    & 0.05    & 0.03     & 0.05  \\ 
\noalign{\hrule height 0.5pt}
\end{tabular}
\label{tab:occ_Be10}
\end{table}

In Fig.~\ref{fig:Be10_AV_MN}, we show the energy spectrum of $^{10}$Be using MN interaction which does not have the tensor force. 
The whole trend of the spectrum agrees with the experiment for low-lying states except for the $0^+_2$, 
which is highly located at $E_x$=16 MeV in TOSM with MN interaction.
In the $0^+_2$ state, two neutrons are excited from $0p_{3/2}$ to $0p_{1/2}$, which makes the large splitting energy.
This is related to the large splitting energy of $^9$Be between the $\frac32^-_1$ and $\frac12^-_1$ as shown in Fig.~\ref{fig:MN_Be9}.

In Table \ref{tab:10Be_ham}, we show the Hamiltonian component using AV8$^\prime_{\rm eff}$.
Among the $0^+$ states, the ground $0^+$ state has the largest contribution of the kinetic part and also of the central and $LS$ force.
On the other hand, the $0^+_2$ state has the largest tensor contribution.
If the $0^+_2$ state has a relation to the developed $\alpha$ cluster state, there might appear the strong tensor contribution from the $\alpha$ clusters in the nucleus.
The results of TOSM might suggest the $\alpha$ clustering in the $0^+_2$ of $^{10}$Be.

\begin{table}[t]
\centering
\caption{Hamiltonian components for $^{10}$Be with $T$=1 in MeV.}
\begin{tabular}{c|ccccc}  
\noalign{\hrule height 0.5pt}
State   &  Energy  &  Kinetic    &  Central & Tensor    &  $LS$   \\
\noalign{\hrule height 0.5pt}
$0^+_1$ & $-29.91$  & $245.17$  & $-146.18$ & $-108.78$  & $-20.13$ \\
$0^+_2$ & $-21.98$  & $239.92$  & $-136.77$ & $-109.86$  & $-15.27$ \\
$0^+_3$ & $-18.00$  & $238.28$  & $-138.19$ & $-105.00$  & $-13.09$ \\
$0^+_4$ & $-14.93$  & $235.56$  & $-126.81$ & $-106.01$  & $-17.66$ \\
\noalign{\hrule height 0.5pt}
$1^+_1$ & $-24.09$  & $240.83$  & $-136.22$ & $-110.68$  & $-18.03$ \\
$1^+_2$ & $-22.08$  & $240.24$  & $-137.64$ & $-107.53$  & $-17.14$ \\
$1^+_3$ & $-17.06$  & $237.27$  & $-128.07$ & $-107.73$  & $-18.53$ \\
$1^+_4$ & $-14.32$  & $234.37$  & $-127.18$ & $-107.43$  & $-14.07$ \\
\noalign{\hrule height 0.5pt}
$2^+_1$  & $-27.30$ & $244.03$ & $-142.77$ &  $-108.07$  & $-20.48$ \\
$2^+_2$  & $-25.15$ & $242.71$ & $-142.04$ &  $-108.49$  & $-17.33$ \\
$2^+_3$  & $-23.46$ & $241.13$ & $-138.09$ &  $-107.58$  & $-18.92$ \\
$2^+_4$  & $-20.89$ & $239.81$ & $-136.68$ &  $-107.90$  & $-16.12$ \\
\noalign{\hrule height 0.5pt}
$3^+_1$ & $-22.61$ & $241.53$ & $-139.81$ & $-107.50$ & $-16.82$ \\
$3^+_2$ & $-18.90$ & $238.09$ & $-131.79$ & $-106.81$ & $-18.38$ \\
$3^+_3$ & $-17.29$ & $236.83$ & $-130.74$ & $-107.98$ & $-15.41$ \\
$3^+_4$ & $-16.55$ & $237.11$ & $-132.14$ & $-106.35$ & $-15.17$ \\
\noalign{\hrule height 0.5pt}
$4^+_1$ & $-21.40$ & $240.82$ & $-136.36$ & $-105.72$ & $-20.14$ \\
$4^+_2$ & $-17.55$ & $238.17$ & $-134.68$ & $-105.47$ & $-15.58$ \\
$4^+_3$ & $-15.61$ & $236.28$ & $-131.51$ & $-106.35$ & $-14.03$ \\
\noalign{\hrule height 0.5pt}
\end{tabular}
\label{tab:10Be_ham}
\end{table}

\subsection{$^{10}$B}\label{sec:B10}

\begin{figure}[t]
\centering
\hspace*{-0.5cm}
\includegraphics[width=9.0cm,clip]{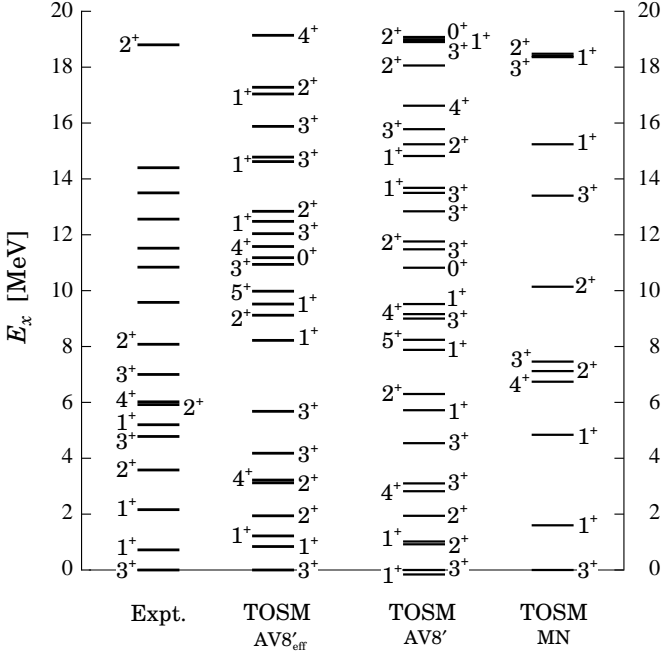}
\caption{Comparison of the excitation energy spectrum of $^{10}$B with $T$=0 using AV8$^\prime_{\rm eff}$, AV8$^\prime$ and MN interactions in TOSM.}
\label{fig:B10_AV_MN}
\end{figure}

We analyze the level structure of $^{10}$B with $T=0$ states as shown in Fig.~\ref{fig:B10_AV_MN}.
The binding energy of $^{10}$Be with AV8$^\prime_{\rm eff}$ is obtained as $29.13$ MeV, which is smaller than the experimental value $64.75$ MeV.
In the figure, we employ the two kinds of $NN$ interaction; one is AV8$^\prime_{\rm eff}$ which is used for $^8$Be, $^9$Be and $^{10}$Be.
For reference, the other is the original AV8$^\prime$ without the modification of the tensor and $LS$ forces.

It is found that the spin of the ground state is obtained as $1^+$ state for AV8$^\prime$, which is different from the experimental situation.
This result is commonly obtained for other calculations using the bare $NN$ interaction without the three-nucleon interaction \cite{pieper01}.
On the other hand, in the case of AV8$^\prime_{\rm eff}$, we can reproduce the ground state spin and also the low-lying spectra.
In particular, the number of levels for each spin are almost reproduced in TOSM.
These results indicate that the effective treatment of the tensor and $LS$ forces gives the proper state-dependence to explain the level order of $^{10}$B,
although this treatment is not related to the three-nucleon interaction.
The matter radius of the ground state is obtained as 2.20 fm as shown in Table \ref{tab:radius}.

\begin{table}[t]
\centering
\caption{Hamiltonian components for $^{10}$B with $T=0$ in MeV.}
\begin{tabular}{c|ccccc}  
\noalign{\hrule height 0.5pt}
State   &  Energy  &  Kinetic   &  Central  & Tensor     &  $LS$   \\
\noalign{\hrule height 0.5pt}
$0^+_1$ & $-17.95$  & $237.52$  & $-124.77$ & $-110.81$  & $-19.89$ \\
$0^+_2$ &  $-6.34$  & $230.69$  & $-122.20$ & $-106.50$  & $ -8.33$ \\
\noalign{\hrule height 0.5pt}
$1^+_1$ & $-28.30$  & $224.92$  & $-142.09$ & $-113.98$  & $-17.15$ \\
$1^+_2$ & $-27.90$  & $244.67$  & $-140.88$ & $-112.57$  & $-19.17$ \\
$1^+_3$ & $-20.92$  & $240.35$  & $-138.94$ & $-108.90$  & $-13.42$ \\
$1^+_4$ & $-19.61$  & $239.07$  & $-134.93$ & $-108.84$  & $-14.90$ \\
\noalign{\hrule height 0.5pt}
$2^+_1$ & $-27.20$  & $243.81$  & $-141.27$ & $-113.08$  & $-16.66$ \\
$2^+_2$ & $-26.01$  & $243.19$  & $-140.16$ & $-113.27$  & $-15.77$ \\
$2^+_3$ & $-20.01$  & $239.29$  & $-137.94$ & $-110.68$  & $-10.68$ \\
$2^+_4$ & $-16.29$  & $236.51$  & $-126.45$ & $-109.25$  & $-17.11$ \\
\noalign{\hrule height 0.5pt}
$3^+_1$ & $-29.13$  & $245.40$  & $-141.60$ & $-111.81$ & $-21.12$ \\
$3^+_2$ & $-24.96$  & $242.99$  & $-139.21$ & $-111.33$ & $-17.47$ \\
$3^+_3$ & $-23.44$  & $241.30$  & $-136.57$ & $-111.69$ & $-16.47$ \\
$3^+_4$ & $-18.18$  & $238.73$  & $-134.21$ & $-107.38$ & $-15.34$ \\
\noalign{\hrule height 0.5pt}
$4^+_1$ & $-25.93$  & $243.42$  & $-138.77$ & $-110.92$ & $-19.65$ \\
$4^+_2$ & $-17.42$  & $237.81$  & $-133.97$ & $-109.02$ & $-12.23$ \\
$4^+_3$ & $-10.00$  & $233.34$  & $-124.36$ & $-106.86$ & $-12.10$ \\
\noalign{\hrule height 0.5pt}
$5^+_1$ & $-19.23$  & $239.95$  & $-135.30$ & $-108.49$ & $-15.40$ \\
\noalign{\hrule height 0.5pt}
\end{tabular}
\label{tab:10B_ham}
\end{table}

We also calculate the $^{10}$B energy spectrum by using the MN interaction for comparison.
The result is shown in Fig.~\ref{fig:B10_AV_MN}. It is found that level density is smaller than the experiment.
One of the reasons of this result comes from the strong effect of the effective $LS$ force in MN for $^{10}$B as similarly seen in $^9$Be and $^{10}$Be. 
The ground state radius of $^{10}$B is obtained as 1.88 fm. This small radius in the MN interaction affects the saturation property and provides the small level density.

In Table \ref{tab:10B_ham}, we list the Hamiltonian components for each state using AV8$^\prime_{\rm eff}$. 
In the ground state region, the $1^+_1$ state shows the largest tensor contribution and also the largest kinetic energy, which are correlated each other by the tensor force. 
The $3^+_1$ shows the largest $LS$ contribution. The $2^+_{1,2}$ states also show the rather large tensor contribution.

Comparing $^{10}$B and $^{10}$Be in the Hamiltonian components, 
it is found that the tensor contributions are rather larger in $^{10}$B than those of $^{10}$Be.
This trend is natural from the view point of the attractive effect of $T=0$ channel of the tensor force.
On the other hand, the $LS$ contributions entirely do not show the large difference in two nuclei.

\section{Summary}\label{sec:summary}

The nucleon-nucleon ($NN$) interaction has two specific characters, the tensor force originated from the pion-exchange and the short-range repulsion.
We describe these two characters of the $NN$ interaction in nuclei on the basis of the tensor-optimized shell model (TOSM) with unitary correlation operator method (UCOM)
, TOSM+UCOM.
The TOSM basis states optimize the two-particle two-hole (2p2h) states fully by using the Gaussian expansion method.
The 2p2h states in TOSM play important role for the description of strong tensor correlation with high-momentum of nucleon motion in nuclei.
Using TOSM+UCOM, we have analyzed three nuclei, $^{9,10}$Be and $^{10}$B as the extension of the previous analysis of $^8$Be.
In this paper, we mainly investigated the structure difference between the low-lying and the excited states of $^9$Be with two isospin $T$=1/2 and $T$=3/2 states.
We used the effective $NN$ interaction based on the AV8$^\prime$ interaction, 
which retains the characteristics of the bare $NN$ interaction and simulates the few-body $^4$He calculation.

For $^9$Be, it is found that TOSM nicely reproduces the excitation energy spectrum of $^9$Be for two isospin states, 
except for the small energy distance between the low-lying states and the highly excited states, the latter group is close to the $T$=3/2 states in the excitation energy.
The small energy distance is considered to come from the missing $\alpha$ cluster component in the low-lying states in TOSM.
We have obtained the same situation for $^8$Be with $T$=0 and $T=1$ states.
The common result between $^8$Be and $^9$Be indicates the necessity of the explicit component of $\alpha$ clustering in the TOSM basis states for two nuclei,
in particular, in the low-lying energy region, in which the $\alpha$ cluster correlation is considered to exist strongly.

For highly excited states, the energy spectrum of $^9$Be is normalized to the $\frac32^-$($T$=3/2) state,
because this state can correspond to the isobaric analog state of the $^9$Li ground state, which is successfully described in TOSM.
The normalization of the energy spectrum is useful to find the energy locations of the low-lying states and the highly excited states relatively.
The TOSM is found to give almost correct level order of the experiments for $T$=1/2 and $T$=3/2 states, although the spins of highly excited states are not experimentally confirmed yet.
This result indicates that state-dependence of the $NN$ interaction is correctly treated in TOSM.
We also use the interaction without the tensor force, the Minnesota interaction, which gives a different energy level order.
This difference means that the state-dependence of the $NN$ tensor force works important to explain the level order of $^9$Be.
It is found that the $T$=1/2 states of $^9$Be have the stronger tensor contribution than those of the $T$=3/2 states.
This can be understood from the $T$=0 attractive channel of the tensor force, originated from the one-pion exchange phenomenon.

We have also investigated the dependence of the tensor matrix elements on the $^9$Be states to see the effect of tensor force explicitly.
The tensor force gives the larger attraction for $T$=1/2 states than for $T$=3/2 ones for $^9$Be, which makes the energy difference of two isospin states large. 
This results is also confirmed in the $^8$Be analysis between $T$=0 and $T$=1 states.

For $^{10}$Be, we have obtained the nice energy spectrum for low-lying states, while the highly excited states are not confirmed for spins experimentally. 
Among the $0^+$ states, $0^+_2$ state possesses the largest tensor contribution and is dominated by the $p$-shell configuration and the mixing of $sd$-orbit is small.
This situation is different from the recent theoretical analysis of $^{10}$Be ($0^+_2$) by using the $\alpha$ cluster model.
The $\alpha$ cluster model suggest the large mixing of $sd$-orbit of valence two neutron with the developed two-$\alpha$ clustering.
In TOSM, the small mixing of $sd$-orbit is also related to the higher energy of $\frac12^+$ state of $^9$Be.
It is also found that the tensor contribution of $0^+_2$ is the largest value among the $0^+$ states. This might be related to the $\alpha$ clustering in this state.

For $^{10}$B, we have reproduced the correct spin of the ground state using the effective $NN$ interaction.
The tensor contributions of each state of $^{10}$B are generally larger than those of the $^{10}$Be.
This is because the number of $pn$ pair is larger in $^{10}$B than $^{10}$Be, which plays an important role 
on the tensor correlation in nuclei. 

For the interaction, we phenomenologically introduce the effective $NN$ interaction based on the bare interaction for TOSM, 
which entirely describes the level order of $^{9,10}$Be and $^{10}$B in addition to the results of $^8$Be.
It is interesting to examine the applicability of this interaction to the systematic description of the light nuclei in the future.

\section*{Acknowledgment}
We would like to thank Professor Hisashi Horiuchi for his continuous encouragement and fruitful discussions on the tensor correlation in nuclei.
This work was supported by JSPS KAKENHI Grant Numbers, 24740175, 23224004 and 15K05091.
Numerical calculations were performed on a computer system at Research Center for Nuclear Physics, Osaka University.

\def\JL#1#2#3#4{ {{\rm #1}} \textbf{#2}, #4 (#3)}  
\nc{\PR}[3]     {\JL{Phys. Rev.}{#1}{#2}{#3}}
\nc{\PRC}[3]    {\JL{Phys. Rev.~C}{#1}{#2}{#3}}
\nc{\PRA}[3]    {\JL{Phys. Rev.~A}{#1}{#2}{#3}}
\nc{\PRL}[3]    {\JL{Phys. Rev. Lett.}{#1}{#2}{#3}}
\nc{\NP}[3]     {\JL{Nucl. Phys.}{#1}{#2}{#3}}
\nc{\NPA}[3]    {\JL{Nucl. Phys.}{A#1}{#2}{#3}}
\nc{\PL}[3]     {\JL{Phys. Lett.}{#1}{#2}{#3}}
\nc{\PLB}[3]    {\JL{Phys. Lett.~B}{#1}{#2}{#3}}
\nc{\PTP}[3]    {\JL{Prog. Theor. Phys.}{#1}{#2}{#3}}
\nc{\PTPS}[3]   {\JL{Prog. Theor. Phys. Suppl.}{#1}{#2}{#3}}
\nc{\PTEP}[3]   {\JL{Prog. Theor. Exp. Phys.}{#1}{#2}{#3}}
\nc{\PRep}[3]   {\JL{Phys. Rep.}{#1}{#2}{#3}}
\nc{\AP}[3]     {\JL{Ann. Phys.}{#1}{#2}{#3}}
\nc{\JP}[3]     {\JL{J. of Phys.}{#1}{#2}{#3}}
\nc{\andvol}[3] {{\it ibid.}\JL{}{#1}{#2}{#3}}
\nc{\PPNP}[3]   {\JL{Prog. Part. Nucl. Phys.}{#1}{#2}{#3}}
\nc{\FBS}[3]   {\JL{Few Body Syst.}{#1}{#2}{#3}}

\end{document}